\documentclass[11pt]{article}
\pdfoutput=1
\usepackage{jheppub}
\usepackage{amsfonts}
\usepackage{amsmath}
\usepackage{mathtools}
\allowdisplaybreaks[4]         
\usepackage{amssymb}   
\usepackage{euscript}       
\usepackage[dvipsnames]{xcolor}
\usepackage{upgreek}
\usepackage{mathrsfs}
\usepackage{graphicx}
\usepackage{caption}
\usepackage{subcaption}
\usepackage[export]{adjustbox}
\usepackage[makeroom]{cancel}
\usepackage{physics}
\usepackage{tensor}
\usepackage{braket}
\usepackage{float}
\usepackage{color}
\usepackage[normalem]{ulem}
\usepackage{starfont}

\usepackage{hyperref}
\hypersetup{
	colorlinks =WildStrawberry,
	linkcolor =red,
	pdftitle={},
	citecolor=MidnightBlue,
	filecolor=WildStrawberry,
	urlcolor=WildStrawberry,
}

\newcommand{\rh}{r_\text{h}}
\newcommand{\req}[1]{(\ref{#1})} 

\newcommand{\be}{\begin{equation}}
\newcommand{\ee}{\end{equation}}

\newcommand{\X}{\Phi_0}
\newcommand{\R}{\mathcal{R}}
\newcommand{\E}{\mathcal{E}}
\newcommand{\diff}{\mathrm{d}}
\newcommand{\G}{G_\mathrm{N}}

\newcommand{\eg}{{\it e.g.,}\ }
\newcommand{\ie}{{\it i.e.,}\ }

\makeatletter
\newcommand{\dal}{\mathop{\mathpalette\dal@\relax}}
\newcommand{\dal@}[2]{%
  \begingroup
  \sbox\z@{$\m@th#1\square$}%
  \dimen0=\fontdimen8
    \ifx#1\displaystyle\textfont\else
    \ifx#1\textstyle\textfont\else
    \ifx#1\scriptstyle\scriptfont\else
    \scriptscriptfont\fi\fi\fi3
  \makebox[\wd\z@]{%
    \hbox to \ht\z@{%
      \vrule width \dimen0
      \kern-\dimen0
      \vbox to \ht\z@{
        \hrule height \dimen0 width \ht\z@
        \vss
        \hrule height 2\dimen0
      }%
      \kern-2.5\dimen0
      \vrule width 2.5\dimen0
    }%
  }%
  \endgroup
}
\makeatother

\title{\huge Electromagnetic Generalized Quasi-topological gravities in $(2+1)$ dimensions}

\author[\text{\Zeus}]{Pablo Bueno,} 
\author[\text{\Kronos}]{Pablo A. Cano,} 
\author[\text{\Hades},\text{\Vulkanus}]{Javier Moreno}
\author[\text{\Poseidon}]{and Guido van der Velde}

\affiliation[\text{\Zeus}]{Departament de F\'isica Qu\`antica i Astrof\'isica, Institut de Ci\`encies del Cosmos, \\
Universitat de Barcelona, Mart\'i i Franqu\`es 1, E-08028 Barcelona, Spain.}

\affiliation[\text{\Kronos}]{Instituut voor Theoretische Fysica, KU Leuven,\\
 Celestijnenlaan 200D, B-3001 Leuven, Belgium.}

\affiliation[\text{\Hades}]{Department of Physics and Haifa Research Center for Theoretical Physics and Astrophysics,\\ University of Haifa, Haifa 31905, Israel}

\affiliation[\text{\Vulkanus}]{Department of Physics, Technion, Israel Institute of Technology,\\
Haifa, 32000, Israel.}

\affiliation[\text{\Poseidon}]{Instituto Balseiro, Centro At\'omico Bariloche,\\
 8400-S.C. de Bariloche, R\'io Negro, Argentina.}

\vspace{0.3cm}
\emailAdd{pablobueno@ub.edu, pabloantonio.cano@kuleuven.be, jmoreno@campus.haifa.ac.il, guido.vandervelde@ib.edu.ar} 

\abstract{The construction of Quasi-topological gravities in three-dimensions requires coupling a scalar field to the metric. As shown in \href{http://www.arXiv.org/abs/2104.10172}{{\tt arXiv:2104.10172}}, the resulting ``Electromagnetic'' Quasi-topological (EQT) theories admit  charged black hole solutions characterized by a single-function for the metric, $-g_{tt}=g^{-1}_{rr}\equiv f(r)$,  and a simple azimuthal form for the scalar. Such black holes, whose metric can be determined fully analytically, generalize the BTZ solution in various ways, including singularity-free black holes without any fine-tuning of couplings or parameters. In this paper we extend the family of EQT theories to general curvature orders. We show that, beyond linear order, $f(r)$ satisfies a second-order differential equation rather than an algebraic one, making the corresponding theories belong to the Electromagnetic Generalized Quasi-topological (EGQT) class. We prove that at each curvature order, the most general EGQT density is given by a single term which contributes nontrivially to the equation of $f(r)$ plus densities which do not contribute at all to such equation. The proof relies on the counting of the exact number of independent order-$n$ densities of the form $\mathcal{L}(R_{ab},\partial_a \phi)$, which we carry out. We study some general aspects of the new families of EGQT black-hole solutions, including their thermodynamic properties and the fulfillment of the first law, and explicitly construct a few of them numerically.

}

\begin{document}

 	\maketitle  

\section{Introduction}
Gravity is different in three spacetime dimensions. On the one hand, the three-dimensional Riemann tensor is fully determined by the Ricci tensor. This implies that all Einstein metrics are locally equivalent to maximally symmetric spacetimes and that ---in the absence of additional fields--- no local degrees of freedom propagate \cite{Deser:1983tn}. Classical three-dimensional Einstein gravity is far from trivial though. In the presence of a negative cosmological constant this is reflected, for instance, in the existence of black hole solutions \cite{Banados:1992wn,Banados:1992gq}. In spite of various fundamental differences with its higher-dimensional cousins ---such as being locally equivalent to pure AdS$_3$ and possessing no curvature singularity--- the BTZ black hole does feature many of their characteristic properties, including the presence of event and Cauchy  horizons, or their thermodynamic and holographic interpretations.

Since all non-Riemann curvatures can be generically removed from the action by field redefinitions, all higher-curvature corrections to Einstein gravity built  exclusively from the metric are trivial when considered as perturbative corrections. However, when included with finite couplings, they typically give rise to nontrivial local dynamics ---manifest in the propagation of massive spin-2 and spin-0 modes \cite{Gullu:2010sd,Bueno:2022lhf}. Also, while the BTZ black hole is a solution of all higher-curvature gravities, new solutions have also been constructed in certain cases. For instance, New Massive Gravity \cite{Bergshoeff:2009hq} and its  generalizations \cite{Gullu:2010pc, Sinha:2010ai,Paulos:2010ke} allow for black holes which differ from the BTZ by: being  locally inequivalent  to  AdS$_3$; possessing  dS$_3$, flat, or  Lifshitz  asymptotes; or presenting curvature singularities \cite{Bergshoeff:2009aq,Oliva:2009ip,Clement:2009gq,Alkac:2016xlr,Barnich:2015dvt,AyonBeato:2009nh,Gabadadze:2012xv,Ayon-Beato:2014wla,Fareghbal:2014kfa,Nam:2010dd,Gurses:2019wpb,Bravo-Gaete:2020ftn}. Various classes of new black holes arise when gravity is coupled to additional fields. These include solutions to Einstein-Maxwell \cite{Clement:1993kc,Kamata:1995zu,Martinez:1999qi,  Hirschmann:1995he, Cataldo:1996yr, Dias:2002ps, Cataldo:2004uw,  Cataldo:2002fh},   Einstein-Maxwell-dilaton \cite{Chan:1994qa, Fernando:1999bh, Chen:1998sa,Koikawa:1997am,Edery:2020kof,Edery:2022crs} and  Maxwell-Brans-Dicke \cite{Sa:1995vs,Dias:2001xt} systems. For these, the metric typically presents a curvature singularity and some of the matter fields include logarithmic profiles. Black hole solutions of various types have also been found  for theories with minimally  and non-minimally coupled scalar fields \cite{Martinez:1996gn,Henneaux:2002wm,Correa:2011dt,Zhao:2013isa,Tang:2019jkn,Karakasis:2021lnq,Baake:2020tgk,Nashed:2021jvn,Karakasis:2021ttn,Arias:2022jax,Desa:2022gtw}. Some of these are obtained from well-defined limits of Lovelock gravities to  three-dimensions,  \cite{Hennigar:2020drx,Hennigar:2020fkv,Ma:2020ufk,Konoplya:2020ibi,Lu:2020iav}. 
Again, the solutions generically contain  curvature singularities and, sometimes, globally regular scalars.
There also exist black-hole solutions to gravity theories coupled to non-linear electrodynamics  in  various  ways \cite{Cataldo:1999wr,Myung:2008kd,Mazharimousavi:2011nd,Mazharimousavi:2014vza,Hendi:2017mgb,Guerrero:2021avm,Maluf:2022jjc}. In some cases, special choices of the modified electromagnetic Lagrangian give rise to singularity-free black holes  \cite{Cataldo:2000ns,He:2017ujy,HabibMazharimousavi:2011gh}.  On a different front, in the context of braneworld holography, modified versions of the BTZ black hole which incorporate the full backreaction of strongly coupled quantum fields have been studied in \cite{Emparan:2002px,Emparan:2020znc,Emparan:2022ijy}.

In addition to these results, a new collection of analytic generalizations of the BTZ black hole was presented in \cite{Bueno:2021krl}. The new configurations solve the equations of a new family of Einstein gravity modifications which involve a non-minimally coupled scalar field and belong to the so-called ``Electromagnetic Quasi-topological'' (EQT) class ---see \req{eq:EQT} below for the action. The solutions take the form 
\begin{equation}\label{eq:SSSM}
\diff s^2=-f(r)\diff t^2+\frac{\diff r^2}{f(r)}+r^2\diff\varphi \, ,\quad \phi=p\varphi\, , 
\end{equation}
with  $r\geq 0$, $\varphi=[0,2\pi)$, and where $p$ is an arbitrary dimensionless constant. It is notable the presence of a single function, $f(r)$, characterizing the metric, as well as the explicit ``magnetic'' form of the scalar field.\footnote{The scalar field should be understood in this context as a $0$-form with field strength $\diff \phi$, which is the gauge-invariant quantity. Analogously to a magnetic monopole in $D=4$, the scalar field $\phi$ can be defined globally by gluing two patches.} The general solution ---see \req{rfs}--- includes various types of spacetimes, depending on the values of the gravitational couplings. These include black holes with one or several horizons and with various kinds of possible singularities (BTZ-like, conical and curvature singularities), including completely regular ones. In some of the latter, global regularity is achieved without imposing any kind of fine tuning between the gravitational couplings and the physical parameters of the solutions. This situation occurs for solutions which behave as $f(r)\overset{r\rightarrow 0}{\longrightarrow} \mathcal{O}(r^{2s})$ with $ s>1$ and is special to three-dimensions.

The results in \cite{Bueno:2021krl} were motivated by a set of ideas which have led to the construction of numerous higher-curvature generalizations of Einstein gravity black holes in higher dimensions \cite{Oliva:2010eb,Myers:2010ru,Dehghani:2011vu,Bueno:2016xff,Hennigar:2016gkm,Bueno:2016lrh,Hennigar:2017ego,Hennigar:2017umz,Bueno:2017sui,Ahmed:2017jod,Feng:2017tev,Bueno:2017qce,Bueno:2019ycr,Cisterna:2017umf,Arciniega:2018fxj,Cisterna:2018tgx,Bueno:2019ltp,Cano:2020ezi,Cano:2020qhy,Frassino:2020zuv,KordZangeneh:2020qeg,Bueno:2022res}. The corresponding theories possess static solutions which are continuous deformations of the Schwarzschild black hole. These are characteristically determined by a single metric function $f(r)$ which: either satisfies an algebraic equation (in that case, the theories are called ``Quasi-topological'' (QT) gravities) or, alternatively,  satisfies a second order differential equation (in that case, the theories are called ``Generalized Quasi-topological'' (GQT) gravities instead\footnote{Often, the term ``Generalized Quasi-topological'' is used to refer to both types when considered together.}). As argued in \cite{Cano:2020ezi,Cano:2020qhy} and later in \cite{Bueno:2021krl,Cano:2022ord}, matter fields can be implemented within this framework in a natural way. Explicitly, the idea is to non-minimally couple gravity to a $(D-2)$-form field-strength $H$ and consider solutions for which $H $ is given by the volume form of the space transverse to the $(t,r)$ directions in the black hole ansatz times an arbitrary constant $p$. This ``magnetic'' ansatz automatically solves the equation of motion of $H$. For instance, in four dimensions, this corresponds to a standard $2$-form field strength given by $H= p \sin \theta \diff \theta \wedge \diff \phi$.  While $H$ is automatically determined for all theories in a unique way, its presence changes the metric in a nontrivial way. In the case of interest here, the $1$-form field strength is given by $\diff \phi$, where $\phi$ is a real scalar, and the magnetic ansatz is just $\diff \phi = p\, \diff \varphi$, as shown in \req{eq:SSSM}. The presence of $H$ in the corresponding Lagrangians gives rise to the notion of ``Electromagnetic Generalized Quasi-topological'' (EGQT) theories, in order to distinguish them from the purely gravitational Generalized Quasi-topological ones.\footnote{EGQT gravities should not be confused with the ``Quasi-topological Electromagnetism'' theories studied \eg in \cite{Liu:2019rib,Cisterna:2020rkc,Cisterna:2021ckn,Li:2022vcd}.} 

 In the three-dimensional case, it was shown in \cite{Bueno:2022lhf} that only trivial GQT theories exist in the absence of non-gravitational fields. As mentioned above, the situation changes when a scalar field is introduced. We say that a three-dimensional Lagrangian  $\mathcal{L}[R_{ab},\partial_a\phi]$ belongs to the EGQT class when the reduced Lagrangian obtained from evaluating $\sqrt{|g|}\mathcal{L}_{\rm EGQT}$ on the single-function magnetic ansatz \req{eq:SSSM} becomes a total derivative. 
In that case, the equations of motion of the theory reduce to a single equation for the metric function $f(r)$ \cite{Cano:2022ord,Bueno:2021krl}.   In this paper we characterize all EGQT gravities in three dimensions. In particular, we show that the most general theory of that kind can be written as\footnote{We show this to be true up to densities which make no contribution at all to the equation of $f(r)$. As a consequence, we refer to those as ``trivial'' densities, even though they generally possess nontrivial equations for other backgrounds.}
\begin{equation}\label{eq:PLag0}
I_{\text{EGQT}}=\frac{1}{16\pi G}\int\diff^3x\sqrt{|g|}\left[R+\frac{2}{L^2}+\sum_{k=1}\beta_{0,k} L^{2(k-1)} \X^{k}-\sum_{n=1} \mathcal{G}_n\right],
\end{equation}
where the order-$n$ EGQT family reads
\begin{align}\label{Gn}
\mathcal{G}_n\equiv &\sum_{k=0} \frac{(-1)^{n}\beta_{n,k}}{n} L^{2(k+2n-1)} \X^{ k}\left[(2k+5n-2)\Phi_1-n \X R\right]  \Phi_1 ^{n-1}\,,  
\end{align}
and we defined
\begin{equation}
\X \equiv g^{ab}\partial_{a}\phi\partial_b \phi\, , \quad \Phi_1\equiv R_{ab}\partial^a\phi\partial^b\phi\, .
\end{equation}
In these expressions, $L$ is a length scale and $\beta_{n,k}$ are dimensionless constants. Observe also that at each curvature order, $\mathcal{G}_n$ contains terms with an arbitrary number of derivatives of the scalar field. For $n=1$, the above action reduces to the one presented in \cite{Bueno:2021krl}, and we find that it is only in that case that the equation for $f(r)$ is algebraic. Indeed, the equations of motion for the action \req{eq:PLag0} admit solutions of the form \req{eq:SSSM} where $f(r)$ satisfies
\begin{equation}\label{EoM}
\frac{r^2}{L^2}-f-\lambda-\beta_{0,1} p^2 \log \left[\frac{r}{L}\right]+\sum_{k=2} \frac{\beta_{0,k}  p^{2k}  L^{2(k-1)}}{2(k-1) r^{2(k-1)}}+\sum_{n=1}\E_{(n)}=0\, .
\end{equation} 
Here, $\lambda$ is an integration constant related to the mass of the solution and $\E_{(n)}$ reads
\begin{equation}\label{eomf}
\E_{(n)}=\sum_{k=0}\frac{-\beta_{n,k}L^{2(2n-1+k)} p^{2(k+n)} }{n r^{3 n+2k-1}}\left[n(3n+2k-2)f {f'}^{(n-1)}+(n-1)r [ {f'}^{n}-n  f {f'}^{n-2}f'' ]\right]\, ,
\end{equation}
where it is manifest that only for $n=1$ the theory is of the EQT class. A detailed review of EQT theories and their solutions can be found in Section \ref{EQTs}.

In order to count the number of EGQT families, in Section \ref{countingg} we count the exact number of independent densities constructed from arbitrary contractions of the Ricci tensor and $\partial_a\phi$. We do so both as a function of the total number of derivatives of fields, and as a function of the curvature order (namely, independently of the number of derivatives of $\phi$). In particular, we find that there exist
\begin{equation}
\#(n)=\left\lfloor\frac{(n+1)}{288}\left(9(-1)^n+n^3+17n^2+95n+184\right)+\frac{1}{2}\right\rfloor \, ,
\end{equation}
 three-dimensional densities $\mathcal{L}(R_{ab},\partial_a\phi)$ of curvature-order $n$,
where $\lfloor \cdot \rfloor $ is the floor function. 

We use this result in Section \ref{EGQTS} to prove that there exists one and only one nontrivial EGQT family of order-$n$ densities, parametrized by an arbitrary function of  $g^{ab}\partial_{a}\phi\partial_b \phi$ (which is obviously order-$0$ in curvature) as well as $\#(n)-n-1$ ``trivial'' densities. We find a recursive formula which allows one to construct arbitrary-order densities from lower-order ones as well as the explicit expression shown in \req{Gn}. Then, we study the near-horizon and asymptotic behavior of the black-hole solutions, finding that there will in general be a unique solution in each case. In order to actually verify their existence, we construct numerically the corresponding solutions in a few cases. 

 We wrap up in Section \ref{conclu}, where we mention a few ideas for the future. Finally, appendix \ref{numbercon} contains a proof that the number of conditions one needs to impose to a general order-$n$ density in order for it to be of the EGQT class equals $n$.

\section{EQT  gravities in three dimensions}\label{EQTs}
In this section we review the Electromagnetic Quasi-topological gravity theories constructed in  \cite{Bueno:2021krl},  which we aim to generalize in the rest of the paper. In section \ref{bht} we present new results regarding the thermodynamic properties of the EQT black holes. In particular, we compute all the relevant quantities in the most general case, and we verify that they satisfy the first law.

\subsection{Review of the theories and their black holes}
In \cite{Bueno:2021krl} we presented a new family of three-dimensional gravity theories non-minimally coupled to a scalar field admitting new classes of static black-hole solutions. The Lagrangian of such a family can be written as 
\begin{equation}\label{eq:EQT}
\mathcal{L}_{\text{EQT}}(R_{ab},\partial_a\phi)=\frac{1}{16\pi \G}\left[R+\frac{2}{L^2}-\sum_{n=0}^1 \mathcal{G}_{n} \right]\, ,
\end{equation}
where we assumed a negative cosmological constant with length scale $L$ and where
\begin{align}
 \mathcal{G}_0 &\equiv +\sum_{i=1} \beta_{0,i} L^{2(i-1)} \Phi_0^{i}\, , \\
 \mathcal{G}_1 &\equiv -\sum_{j=0} \beta_{1,j} L^{2(j+1)}\Phi_0^{j}\left[ (3+2j) \Phi_1- \Phi_0  R \right],
\end{align}
are the Electromagnetic Quasi-topological densities. Here, $\beta_{0,i},\beta_{1,j}$ are dimensionless constants which parametrize the infinite terms contained in both $ \mathcal{G}_0$ and  $\mathcal{G}_1$.  Indeed, observe that the new densities are, respectively, order-$0$ and order-$1$ in curvature, but both contain terms of arbitrarily high order in derivatives of $\phi$.

For these theories, the effective Lagrangian $L_{\text{EQT}}\equiv \sqrt{|g|}\mathcal{L}_{\text{EQT}}(R_{ab},\partial_a\phi)$ becomes a total derivative when evaluated in the magnetic ansatz \eqref{eq:SSSM}. This implies that they admit solutions characterized by a single function $f(r)$. Indeed, the full non-linear equations of motion reduce to a single independent equation for $f(r)$ which can be integrated once and reads \cite{Bueno:2021krl} 
\begin{equation}\label{rfs}
\frac{r^2}{L^2}-f(r)+\mathcal{E}_{(0)}+\mathcal{E}_{(1)}=\lambda \, ,
\end{equation}
where
\begin{align}
\mathcal{E}_{(0)}&\equiv + \sum_{i=2} \frac{\beta_{0,i} p^2}{2(i-1)} \left(\frac{p L}{r} \right)^{2(i-1)} -\beta_{0,1} p^2 \log\left( \frac{r}{L} \right)\, , \\
\mathcal{E}_{(1)}&\equiv -\sum_{j=0} \beta_{1,j}  (2j+1) \left(\frac{p L}{r} \right)^{2(j+1)} f(r)\, , 
\end{align}
are the contributions from $\mathcal{G}_0$ and $\mathcal{G}_1$, respectively, and  $\lambda$ is an integration constant.

Equation \req{rfs} can then be trivially solved for $f(r)$, and one finds  
\begin{equation}\label{fEQT}
f(r)= \frac{\displaystyle \left[ \frac{r^2}{L^2}-\lambda-\beta_{0,1} p^2 \log\left( \frac{r}{L}\right)+ \sum_{i=2} \frac{\beta_{0,i}  p^2}{2(i-1)}\left(\frac{p L}{r}\right)^{2(i-1)} \right]}{\displaystyle \left[ 1+ \sum_{j=0} \beta_{1,j}(2j+1)\left(\frac{p L}{r}\right)^{2(j+1)} \right]} \, ,
\end{equation}
where $\lambda$ is an integration constant.\footnote{This constant is related to the mass of the solution, but the precise identification would require a careful analysis of the conserved charges in the theory \req{eq:EQT} as the density with $\beta_{1,0}$ seems to yield a nontrivial contribution. Instead, below we obtain the mass from the free energy.} The new metrics represent continuous generalizations of the static BTZ black hole \cite{Banados:1992wn,Banados:1992gq}, which is obtained for $\beta_{0,i}=\beta_{1,j}=0$ $\forall i,j$, as well as of its charged version \cite{Clement:1993kc,Kamata:1995zu,Martinez:1999qi}, which we recover for $\beta_{0,i>1}=\beta_{1,j}=0$ $\forall j$. The charged BTZ black hole is obtained as a solution to Einstein gravity minimally coupled to a scalar field because theories of the type $\mathcal{L}(R_{ab},\partial_a\phi)$ are dual to theories with an electromagnetic field, \ie of the form  $\mathcal{L}(R_{ab},F_{cd} \equiv 2\partial_{[c}A_{d]})$, and the  kinetic term of a free scalar simply gets mapped to the usual Maxwell term \cite{Bueno:2021krl}. In the general case,  the dual field strength is defined by
\begin{equation}
F_{ab}=4\pi G\epsilon_{abc}\frac{\partial \mathcal{L}}{\partial\left(\partial_{c}\phi\right)}\, ,
\end{equation}
and one can in principle invert this relation to find $\partial \phi(F)$ and, from this, the dual Lagrangian, $\mathcal{L}_{\rm dual}(R_{ab}, F_{cd})\equiv \mathcal{L}-\frac{1}{8\pi G}F_{ab}\partial_{c}\phi\epsilon^{abc} $. This is something difficult to do in general but, in the present case, one can consider a perturbative expansion in powers of the length scale $L$, which yields, for $\beta_{0,1} \neq 0$,
\begin{align}
\mathcal{L}_{\text{dual}} =\frac{1}{16\pi G}\left[R+\frac{2}{L^2}+\frac{2}{\beta_{0,1}} F_{ab}F^{ab}  -4 L^2\left[\frac{\beta_{0,2}}{\beta_{0,1}^4}(F_{ab}F^{ab})^2+\frac{3\beta_{1,0}}{\beta_{0,1}^2}\tensor{F}{_{a}^{b}}F^{ac}R_{\langle bc\rangle}\right]\right]+\mathcal{O}(L^4)\, ,
\end{align}
 where $R_{\langle bc\rangle}$ is the traceless part of the  Ricci tensor. Observe the appearance of the usual Maxwell term controlled by $\beta_{0,1}$, as anticipated. While the original ``magnetic'' frame Lagrangian may contain a finite number of terms, the dual one contains infinitely many in general. One can choose to work on either frame, and the magnetically charged solutions of the original frame become electrically charged in the dual one, with a field strength given by
$F_{tr}=-\partial_r A_t$, where  the electrostatic potential reads 
\begin{equation}\label{eq:pot}
\begin{aligned}
A_t=&\frac{1}{2}\left[\beta_{0,1}p \log \frac{r}{r_0}-\sum_{i=2}\frac{i\beta_{0,i} p}{2(i-1)}\left(\frac{Lp}{r}\right)^{2(i-1)}-f'(r)L\sum_{j=0}\beta_{1,j}(j+1)\left(\frac{Lp}{r}\right)^{(2j+1)}+2p\beta_{1,0}\right]\\
&+\Phi \, ,
\end{aligned}
\end{equation}
and $r_0$, an IR cutoff, prevents the potential from diverging within the radius $r_0$. Also, $\Phi$ is an integration constant which corresponds to the potential in the asymptotic region and whose value can be fixed by imposing that $A_t(r)$ vanishes at the horizon of the black hole.
Since the expression of the magnetic-frame Lagrangian is simpler, and so is the form of $\phi$ for the class of solutions we are interested in, from now on we will perform all our calculations in such a frame.

Going back to the solutions, observe that, on general grounds, they will describe black holes whenever the function $f(r)$ has at least one positive root. If $f(r)$ has several positive roots, then the solution possesses several horizons and the largest root represents the event horizon, that we denote by
\begin{equation}\label{rhh}
\rh=\text{max}\left\{ r>0 | f(r)=0 \right\}\, .
\end{equation}

 More generally, as we show  in detail in \cite{Bueno:2021krl}, depending on the values of the gravitational couplings, the new metrics describe different types of solutions, including: black holes with curvature singularities, black holes with conical singularities, black holes with ``BTZ-like'' singularities, regular black holes for which $f(r)\overset{r\rightarrow 0}{\rightarrow} 1$ (which require some degree of fine tuning of the parameters), regular black holes for which $f(r) \overset{r\rightarrow 0}{\rightarrow} 0$ (which do not require any fine tuning), as well as solutions with no horizons which are regular everywhere.


\subsection{Black hole thermodynamics}\label{bht}
Let us now consider the thermodynamic properties of the above black hole solutions.\footnote{A previous study of various thermodynamic aspects of these black holes in several particular cases can be found in \cite{Ditta:2022fjz}.} 

The temperature $T=f'(\rh) / (4\pi)$ can be straightforwardly obtained from the metric function \eqref{fEQT} using the horizon condition $f(\rh)=0$. The result reads
\begin{equation}
T=\frac{1}{4\pi}\left[\frac{2\rh}{L^2}-\sum_{i=1}\frac{\beta_{0,i} p^{2i} L^{2(i-1)}}{\rh^{2i-1}}\right]\cdot \left[\displaystyle 1+ \sum_{j=0}\beta_{1,j}(2j+1)\left(\frac{Lp}{\rh}\right)^{2(j+1)}\right]^{-1}\, .
\end{equation}
Regarding the entropy, we can use Wald's formula \cite{Wald:1993nt,Iyer:1994ys}, given by
\begin{equation}\label{Wald}
S=-2\pi\int_\mathrm{h}\diff \varphi\sqrt{h}\, \epsilon\indices{_a^c}\epsilon_{cb}\frac{\partial \mathcal{L}}{\partial R_{ab}},
\end{equation}
where $\mathrm{h}$ is the horizon surface, $\sqrt{h}$ is the determinant of the induced metric and $\epsilon_{ab}$ is the binormal to the horizon surface, normalized so that  $\epsilon_{ab}\epsilon^{ab}=-2$. We find
\begin{equation}
S=\frac{\pi \rh}{2G}\left[1-\sum_{j=0}\beta_{1,j} \left(\frac{L p}{\rh}\right)^{2(j+1)}\right].
\end{equation}
Note that the higher order terms proportional to $\beta_{1,j}$ lead to corrections to the area law.

The free energy $F$ can be computed from the Euclidean on-shell action $F=\beta I_\text{E}$, where $\beta$ is the inverse of the temperature. The full Euclidean action reads
\begin{equation}
I_\text{E}=-\frac{1}{16\pi G}\int_{\mathcal{M}} \diff^3 x\sqrt{g} \mathcal{L} -\frac{1}{8\pi G}\int_{\partial \mathcal{M}}\diff^2x\sqrt{h}\left(K-\frac{1}{L}\right)\, ,
\end{equation}
where the first term is the bulk action, and the second one includes the Gibbons-Hawking-York boundary term \cite{York:1972sj,Gibbons:1976ue} and a suitable counterterm \cite{deHaro:2000vlm,Balasubramanian:1999re,Emparan:1999pm}. 
We find
\begin{equation}
\begin{split}
F&=\frac{1}{8 G}\left[\lambda+\beta_{0,1} p^2 \log{\frac{r_0}{L}}-2 p^2\beta_{1,0}-\rh f'(\rh)+\sum_{j=0}\beta_{1,j} \frac{(L p)^{2(k+1)}}{\rh^{2k+1}} f'(\rh)\right]\\ 
&=\frac{1}{8 G}\left[\frac{\rh^2}{L^2}-\beta_{0,1} p^2 \log{\frac{\rh}{r_0}}+\sum_{i=2}\frac{\beta_{0,i} p^2}{2(i-1)}\left(\frac{p L}{\rh}\right) ^{2(i-1)}-\rh f'(\rh)-2\beta_{1,0} p^2+\right.\\
&\quad\quad\quad~\left. +\sum_{j=0}\beta_{1,j} \frac{(L p)^{2(k+1)}}{\rh^{2k+1}} f'(\rh)\right],
\end{split}
\end{equation}
where in the second line we wrote $\lambda$ in terms of $\rh$ using that $f(r_h)=0$.
The thermodynamic mass of the black holes $M$  can be then obtained found from the relation $F=M-TS$, this is\footnote{Note that this does not coincide with the ADM mass \cite{Abbott:1981ff,Deser:2002jk} reported in \cite{Bueno:2021krl}, the difference being $3\beta_{1,0} p^2$. We suspect that a more detailed calculation of the latter may make the mismatch disappear.}
\begin{equation}
\begin{split}
M&=\frac{1}{8 G}\left[ \lambda +\beta_{0,1} p^2\log\frac{r_0}{L}-2p^2\beta_{1,0}\right]\\
&=\frac{1}{8 G}\left[\frac{\rh^2}{L^2}+\beta_{0,1} p^2\log\frac{r_0}{\rh}-2p^2\beta_{1,0}+\sum_{i=2}\frac{\beta_{0,i} p^2}{2(i-1)}\left(\frac{p L}{\rh}\right)^{2(i-1)}\right]\, .
\end{split}
\end{equation}
Moreover, the electric charge can be trivially computed in the ``magnetic frame", yielding
\begin{equation}
Q=\frac{1}{4\pi G}\int_{\mathbb{S}^1} \diff\phi=\frac{p}{2 G}\, .
\end{equation} 
Finally, we can give an expression for the potential, that is, the quantity conjugate to the charge. This is defined as $\Phi=\lim_{r\to\infty} A_t$, so using \eqref{eq:pot} we find
\begin{equation}\label{eq:chempot}
\Phi=\frac{1}{2}\left[-\beta_{0,1}p\log\frac{\rh}{r_0}-2p\beta_{1,0}+\sum_{i=2}\frac{i \beta_{0,i} p}{2(i-1)}\left(\frac{L p}{\rh}\right)^{2(i-1)}+4\pi T L\sum_{j=0}\beta_{1,j} (j+1)\left(\frac{L p}{\rh}\right)^{2j+1}\right]\, .
\end{equation} 

With all these ingredients, the first law,
\begin{equation}
\diff F=- S\diff T+\Phi \diff Q\, ,
\end{equation}
can be easily verified to hold.\footnote{Relatedly, one can check that the relations
\begin{equation}
S=-\frac{\partial_{\rh} F}{\partial_{\rh} T},\quad \Phi=\frac{1}{\partial_p Q}\left(\partial_p F-\partial_p T\frac{\partial_{\rh} F}{\partial_{\rh} T}\right)\, ,
\end{equation}
are satisfied.}
Equivalently, we can write it in terms of the mass instead of the free energy as
\begin{equation}
\diff M=T\diff S+\Phi\diff Q\, .
\end{equation}


\section{Counting $\mathcal{L}(R_{ab},\partial_a\phi)$ densities}\label{countingg}
Three-dimensional EGQT gravities are theories built from contractions of the Ricci tensor and $\partial_a\phi$ admitting black hole solutions of the form \req{eq:SSSM}.
In order to characterize them, we can start by counting the number of densities of the form $\mathcal{L}(R_{ab},\partial_a\phi)$ which exist in three dimensions. Out of the Ricci tensor alone, it is possible to see that the most general density one can build is a function of the invariants \cite{Paulos:2010ke,Gurses:2011fv}
\begin{equation}
 R\, , \quad \mathcal{R}_2\equiv R_{ab}R^{ab}\, , \quad \mathcal{R}_3 \equiv R_{a}^b R_b^c R_c^a\, .
\end{equation}
The exact number of order-$n$ densities of the form $\mathcal{L}(R_{ab})$ was computed in \cite{Bueno:2022lhf}, the result being
\begin{equation}
\#(n)=\left\lceil \frac{n}{2}\left(\frac{n}{6}+1 \right)+\epsilon \right\rceil
\end{equation}
where $\lceil x\rceil \equiv {\rm min} \{k \in \mathbb{Z} | k \geq x \} $ is the ceiling function and $\epsilon$ is any positive number $\epsilon \ll 1$. We have, for instance, $\#(4104)=1405621$ densities of order $4104$.

Including the scalar into the game amounts to consider three additional building blocks besides the pure-curvature ones. These are
\begin{equation}
\Phi_0 \equiv g^{ab}\partial_a\phi\partial_b\phi \, , \quad \Phi_1\equiv R_{ab}\partial^a\phi \partial^b\phi \, , \quad \Phi_2\equiv R_{ac}R_b^c\partial^a\phi \partial^b\phi \, .
\end{equation}
Invariants whose curvature order is greater or equal to three, $n\geq3$, can be written in terms of lower-order invariants using Schouten identities, analogously to what happens with pure-gravity densities for $n\geq 4$. Indeed,
\begin{equation}\label{eq:SchRPhi}
\delta_{b_1 \dots b_n}^{a_1\dots a_n} R_{a_1}^{b_1}R_{a_2}^{b_2}\cdots \partial_{a_n}\phi\partial^{b_n}\phi=0\, ,\quad \text{for} \quad n\geq3\, .
\end{equation} 
This greatly reduces the number of independent invariants and allows us to express the most general Lagrangian as
\begin{equation}\label{eq:LgRPhi}
\mathcal{L}(R_{ab},\partial_a\phi)=\sum_{i,j,k,l,m,p}\alpha_{ijklmnp}R^i\mathcal{R}^j_2\mathcal{R}^k_3\Phi_0^l\Phi_1^m\Phi_2^p\, .
\end{equation}
The curvature order of a particular density is given by $n=i+2j+3k+m+2p$ whereas the number of scalar fields is given by  $n_\phi=2(l+m+p)$.

We are interested in characterizing the possible monomials from eq.\,\eqref{eq:LgRPhi} through the number of derivatives $\EuScript{N}=2n+n_\phi=2 (i + 2 j + 3 k + l + 2 m + 3 p)$, either in the metric or in the scalar field. For $\EuScript{N}=2,4,6,8$, the independent densities read
\begin{align}
D_{2} &= \{ R,\,  \Phi_0 \} \, , \\
D_{4} &= R \cdot D_{2}  \cup \{ \Phi_0^2,\, \R_2 , \,\Phi_1 \}  \, , \\
D_6 &= R \cdot D_{4}  \cup  \{ \Phi_0^3 ,\, \R_2\X, \,
\Phi_1\X,\, \R_3, \, \Phi_2\}\, , \\
D_8 &=R \cdot D_{6}\cup  \{\Phi_1\Phi_0^2, \, \Phi_0^4,\, \R_2 \Phi_0^2,\, \R_3\X,\, \Phi_2\X ,\, \R_2^2,\R_2\Phi_1,\,\Phi_1^2 \}\, .
\end{align}
So we have $\#(2)=2$, $\#(4)=5$, $\#(6)=10$ and $\#(8)=18$, respectively.
We can obtain the number of possible invariants with $\EuScript{N}$ derivatives using a generating function of the form
\begin{equation}\label{eq:GenFRPhi}
G(x,y)=\frac{1}{(1-x^2)(1-x^4)(1-x^6)(1-y^2)(1-x^2y^2)(1-x^4 y^2)}\, , 
\end{equation}
where $x$ and $y$ parametrize the number of derivatives in the metric tensor and in the scalar field respectively. The Maclaurin series reads
\begin{align}
G(x,y)=&\underbrace{+1}_{\EuScript{N}=0} + \underbrace{(1\cdot x^2 +1\cdot y^2) }_{\EuScript{N}=2}+ \underbrace{(2 x^4 + 2 x^2 y^2 + y^4)}_{\EuScript{N}=4} + \underbrace{(3 x^6 + 4 x^4 y^2 + 
   2 x^2 y^4 + y^6)}_{\EuScript{N}=6}\notag\\
    &+ \underbrace{(4 x^8 + 6 x^6 y^2 + 5 x^4 y^4 + 2 x^2 y^6 + y^8)}_{\EuScript{N}=8}+\ldots\sim \sum_{\EuScript{N}=0}\#(\EuScript{N})x^{\EuScript{N}}
\end{align}
where the coefficients are to be read as follows: there are two densities with $\EuScript{N}=2$, one purely geometrical and another that only depends on the scalar field; then, there are 5 densities with $\EuScript{N}=4$ derivatives: two that depend only on derivatives of the metric, two mixed and one with only derivatives of the scalar field; and so on and so forth.

Setting $x=y$ in \eqref{eq:GenFRPhi}, we are able to explicitly compute the number of densities with $\EuScript{N}$ number of derivatives
\begin{equation}
\#(\EuScript{N}) =\left\lceil\frac{\EuScript{N}^5}{138240}+\frac{\EuScript{N}^4}{2304}+\frac{101\EuScript{N}^3}{10368}+\frac{29\EuScript{N}^2}{288}+\left[\frac{24299}{51840}+\frac{(-1)^{\frac{\EuScript{N}}{2}}}{128}+\frac{1}{81}\cos\left(\frac{\EuScript{N}\pi}{3}\right)\right]\EuScript{N}+\epsilon\right\rceil,
\end{equation}
where $0<\epsilon\ll 1$. This can be written in a somewhat more explicit fashion as 
\begin{equation}
\#(\EuScript{N})= \begin{cases} \frac{1}{960} \left[60((-)^k-1)+k(131+45(-)^k+2k(360+k(410+9k(20+3k)))) \right] \, ,  \\ \frac{1}{960} \left[-75((-)^k-1)+k(941-45(-)^k+2k(900+k(680+9k(25+3k))))\right]\, ,  \\ 
\frac{(2+k)}{960} \left[435+45(-)^k+2k(4+k)(146+27k(4+k))\right] \, ,\end{cases}
\end{equation}
valid for $\EuScript{N}=6k-4$, $\EuScript{N}=6k-2$ and $\EuScript{N}=6k$, respectively. For large $\EuScript{N}$, the number of densities behaves as
\begin{equation}
\#(\EuScript{N}) \approx \frac{\EuScript{N}^5}{138240}+\frac{\EuScript{N}^4}{2304}+\frac{101\EuScript{N}^3}{10368}+\frac{29\EuScript{N}^2}{288}\, .
\end{equation}

It will also be relevant for our purposes to know how many densities there exist of a  certain curvature order $n$ ---\ie independently of the number of derivatives of $\phi$. From this perspective, the possible densities for the first few orders read
\begin{align}
d_1 &= \{R, \,\Phi_{1}   \} \, , \\
d_2 &= R \cdot d_1 \cup \{\R_2, \, \Phi_{1}^2, \, \Phi_{2} \}\, , \\
d_3&= R \cdot d_2 \cup \{\R_3, \, \Phi_{1}^3\, , \Phi_{2}\Phi_{1}, \R_2 \Phi_{1} \}\, , \\ 
d_4&= R\cdot d_3 \cup \{ \Phi_{1}^4\, , \Phi_{2}\Phi_{1}^2, \R_2 \Phi_{1}^2,\,  \R_2^2 ,\, \Phi_{2}^2,\, \R_2 \Phi_{2}\, , \R_3\Phi_{1}  \}\, .
\end{align}
where it is understood that each density can in fact be multiplied by an arbitrary function of $\Phi_0$, which has $n=0$.
Again, we can find the total number of invariant terms at any given order $n$ following the generating functional procedure. In this case, we have
\begin{equation}
G(x)=\frac{1}{(1-x)^2(1-x^2)^2(1-x^3)}\, ,
\end{equation}
where the $(1-x^1)^2$ comes from the $2$ independent order-$1$ invariants, the $(1-x^2)^2$ comes from the $2$ independent order-$2$ invariants, and the  $(1-x^3)^1$ comes from the single independent order-$3$ one. Expanding, we find
\begin{equation}
G(x)=1+2x+5x^2+9x^3+16x^4+25 x^5+\ldots=\sum_{n=0}\#(n)x^n\, .
\end{equation}
We see that the corresponding coefficients match the number of densities enumerated above for $n=1,2,3,4$, namely, $\#(1)=2$, $\#(2)=5$, $\#(3)=9$, $\#(4)=16$. Similarly, we expect $\#(5)=25$, and so on and so forth. The number of invariants of order $n$ can be written in a closed form as 
\begin{equation}\label{pin}
\#(n)=\left\lfloor\frac{(n+1)}{288}\left[9(-1)^n+n^3+17n^2+95n+184\right]+\frac{1}{2}\right\rfloor \, ,
\end{equation}
where $\lfloor x\rfloor \equiv {\rm max} \{k \in \mathbb{Z} | k \leq x \} $ is the floor function. For large $n$, the number of densities grows as 
\begin{equation}
\#(n) \approx \frac{n^4}{288}+\frac{n^3}{16}+\frac{7n^2}{18}\, .
\end{equation}
Again, we stress that each density of order $n$ can be multiplied by an arbitrary function of $\Phi_0$, which means that, strictly speaking, there are infinitely many independent densities at each order $n$. In spite of this, we still find it more illuminating to classify our families of densities as a function of $n$ instead of $\EuScript{N}$.

\section{EGQT gravities in three dimensions}\label{EGQTS}
In this section we construct new families of Electromagnetic Generalized Quasi-topological gravities. We show that there exists exactly one single nontrivial family of theories of that kind at each curvature order $n$ ---each family parametrized by an arbitrary function of $\partial_a\phi \partial^a\phi $. We find a recurrence relation which allows for the construction of general-order densities starting from lower-order ones,  and find an explicit expression for the general-$n$ family. For $n=1$, such family reduces to the Electromagnetic Quasi-topological gravities presented in \cite{Bueno:2021krl} and reviewed in Section \ref{EQTs}, for which the metric function $f(r)$ satisfies an algebraic equation. For $n\geq 2$, we find that the resulting theories are all genuinely ``Generalized'', namely, $f(r)$ satisfies a second-order differential equation instead. We study the near-horizon and asymptotic behavior of the black hole solutions of these theories and construct their profiles numerically in a few cases. Finally, we compute the relevant thermodynamic quantities and verify the first law in the general case.

\subsection{One EGQT family at each order}

When gravity is coupled to a scalar field, there are infinitely many theories which belong to the EGQT class. In fact, a family of EGQT gravities linear in curvature, \ie with $n=1$ ---but involving terms with arbitrarily high $\EuScript{N}$--- was presented in \cite{Bueno:2021krl}. The Lagrangian of such family can be written as
\begin{equation}\label{gg1}
\mathcal{G}_1=\Phi_1 \left(3F_{(1)}\left[\Phi_0\right]+2\X F_{(1)}'\left[\Phi_0\right]\right)- \X R\, F_{(1)}\left[\Phi_0\right] ,
\end{equation}
where $F_{(1)}=F_{(1)}\left[\Phi_0\right]$ is some arbitrary function of the kinetic term of the scalar field $\phi$ and $F_{(1)}'\equiv \frac{dF_{(1)}[x]}{dx}$. To this we can add an additional order-$0$ term involving an arbitrary function of $\Phi_0$, namely,
\begin{equation}\label{g00}
\mathcal{G}_0 = F_{(0)}\left[\Phi_0\right]\, .
\end{equation}
 Writing $F_{(0)}$ and $F_{(1)}$ as series expansions in $\Phi_0$, $\mathcal{G}_0+\mathcal{G}_1$ becomes
\begin{equation}
\mathcal{G}_0+\mathcal{G}_1=\sum_{i=1} \beta_{0,i} L^{2(i-1)} \Phi_0^{i}-\sum_{j=0} \beta_{1,j} L^{2(j+1)} \Phi_0^j \cdot \left[(3+2j) \Phi_1-\Phi_0 R \right]\, ,
\end{equation}
where $\beta_{0,i},\beta_{1,j}$ are arbitrary dimensionless constants, and $L$ is some length scale. The above is the form originally presented in \cite{Bueno:2021krl} and the one we used in Section \ref{EQTs}. 

In this section, we extend the catalog of EGQT densities to arbitrary orders in curvature. The procedure is fairly simple: we consider the most general Lagrangian at a given order $n$, which will be a linear combination of the independent densities, each multiplied by an arbitrary analytic function of the scalar field $F_{i_n,(n)}$. Then, we impose the condition that the reduced Lagrangian $L_f$ becomes a total derivative when evaluated on \req{eq:SSSM}. As a consequence, the functions $F_{i_n,(n)}$ must satisfy a set of relations. Although the number of those is smaller than the number of independent densities, we find that, at each order $n$, there is a single way in which EGQT densities modify the equations of the metric function $f(r)$, namely, we can write the most general order-$n$ EGQT density as a single density which contributes nontrivially to the equation of $f(r)$, plus a sum of densities which make no contribution whatsoever to the equation of $f(r)$ ---sometimes we will call such densities ``trivial'' even though they will not be trivial when evaluated on other backgrounds.

Let us illustrate this with the example of $n=2$. We start considering the most general second-order Lagrangian of the type $\mathcal{L}\left(R_{ab},\partial_a\phi\right)$, this is
\begin{equation}
\mathcal{L}^{(2)}_{\rm general}=F_{1,(2)}R^2+F_{2,(2)}\mathcal{R}_2+F_{3,(2)}R \Phi_1+F_{4,(2)}\Phi_1^2+F_{5,(2)}\Phi_2\,.
\end{equation} 
Now, introducing this expression into the EGQT condition we obtain relations between the free functions, reducing their number to three. The resulting Lagrangian reads
\begin{align}\notag
\tilde{\mathcal{G}}_2=&\mathcal{G}_2+\mathcal{G}_{2,{\rm trivial}}
\end{align}
where 
\begin{align}\label{gg2}
\mathcal{G}_2\ &\equiv F_{(2)}[\Phi_0]\X\left(\frac{R^2}{2}-\mathcal{R}_2\right)-\left(\X {F_{(2)}}'[\Phi_0]+\frac{3}{2}F_{(2)}[\Phi_0]\right)\Phi_2
\end{align}
is a density which contributes nontrivially to the equation of $f(r)$. Also,
\begin{align}
\mathcal{G}_{2,{\rm trivial}} &\equiv G_{1,(2)}\mathcal{T}_{1,(2)}+3G_{2,(2)}\mathcal{T}_{2,(2)}\, , 
\end{align}
where $G_{1,(2)}$, $G_{2,(2)}$ are arbitrary functions of $\Phi_0$ and
\begin{align}
\mathcal{T}_{1,(2)}\equiv  \Phi_1^2-\X  \Phi_2\, ,\quad
\mathcal{T}_{2,(2)}\equiv 3\Phi_2-\X\left(2\mathcal{R}_2-R^2 \right)-2R\Phi_1\, ,
\end{align}
are two densities which vanish when evaluated in such ansatz, $\mathcal{T}_{1,(2)}|_{\req{eq:SSSM}}=\mathcal{T}_{2,(2)}|_{\req{eq:SSSM}}=0$. The first density, $\mathcal{T}_{1,(2)}$, vanishes not only for static and spherically symmetric  ans\"atze, but for all diagonal metrics. The other, $\mathcal{T}_{2,(2)}$, vanishes when evaluated in \req{eq:SSSM}. Therefore, both densities correspond to ``trivial'' EGQT densities and all higher-curvature Lagrangians constructed from linear combinations of them will be as well. We can then take $\mathcal{G}_2$ as a representative of order-$2$ nontrivial EGQT densities as it will differ from any other EGQT of the same order by a trivial density.  

We can apply the same procedure at third order in curvature. In this case, we start with a Lagrangian consisting of nine densities. Then, applying the GQT condition, we obtain the nontrivial family
\begin{equation}\label{gg3}
\mathcal{G}_3=-2F_{(3)}[\Phi_0]\Phi_1\left(\frac{R^2}{2}-\mathcal{R}_2\right)+\left(F_{(3)}[\Phi_0] \,R+\frac{2}{3}{F_{(3)}}'[\Phi_0]\, \Phi_1\right)\Phi_2\, ,
\end{equation}
plus five trivial combinations, which vanish when evaluated in \req{eq:SSSM}.

Interestingly, we find a pattern satisfied by the $\mathcal{G}_n$ of the first few orders when evaluated on \req{eq:SSSM}. In particular, we observe that they satisfy
\begin{equation}\label{eq:onshn}
\mathcal{G}_n\Big|_{\req{eq:SSSM}}=  \frac{p^2}{r^2} F_{(n)}\left[\frac{p^2}{r^2}\right]\left(\frac{f'}{r}\right)^{(n-1)}f''- \frac{ p^2}{r^2}\left( F_{(n)}\left[\frac{p^2}{r^2}\right]+\frac{2p^2}{nr^2} F_{(n)}'\left[\frac{p^2}{r^2}\right]\right)\left(\frac{f'}{r}\right)^{n}\, ,
\end{equation}
which by itself satisfies the EGQT condition for arbitrary $n$, namely: $\sqrt{-g}\mathcal{G}_n\Big|_{ \eqref{eq:SSSM} }$ is a total derivative,
\begin{equation}\label{eq:onshn2}
\sqrt{-g} \mathcal{G}_n\Big|_{\req{eq:SSSM}}=\frac{p^2}{n} \frac{\diff }{\diff r} \left[  F_{(n)}\left[\frac{p^2}{r^2}\right] \left( \frac{f'}{r} \right)^n \right]\, .
\end{equation}
 We suspect that this on-shell expressions are valid for EGQT densities of arbitrary order. In the particular case in which all the general functions of $\Phi_0$ are equal to each other, $F_{(1)}=F_{(2)}=\cdots=F_{(n)}\equiv F$ we can use \req{eq:onshn} to derive a recurrence relation 
\begin{equation}\label{eq:recurr}
 \mathcal{G}_n=\frac{(n-2)(n-1)}{n \Phi_0^2F'[\Phi_0](n-3)}\left[\mathcal{G}_{n-2}\mathcal{G}_2-\mathcal{G}_{n-1}\mathcal{G}_1\right]\, , 
\end{equation}
valid for $n>3$. This means that using lower-order EGQT densities, we can build arbitrarily-high order ones in a recursive way. For instance, using $\mathcal{G}_1$, $\mathcal{G}_2$, $\mathcal{G}_3$ we can construct a new density, $\mathcal{G}_4$, which is of order $n=4$ and is guaranteed, by virtue of \req{eq:onshn2}, to be of the EGQT class. We can go on and build $\mathcal{G}_n$ for general $n$ in an analogous way. 
The relation \req{eq:recurr} tells us that there exists at least one family of EGQT gravities at each order in curvature and, from \req{eq:onshn}, what its on-shell form in our single-function ansatz is. Observe that once the new density $\mathcal{G}_n$ is constructed, the fact that we had to take the functions $F_{(i)}$ to be equal for the densities of various orders involved  simply suggests that for each $n$ the corresponding nontrivial EGQT density depends on a single arbitrary function of $\Phi_0$, which is precisely what we have observed for the first few orders.
 
 This does not exclude, however, the possibility that additional, inequivalent, EGQT densities exist for $n >3$. We will show now that, as a matter of fact, there are no additional nontrivial densities. First, in appendix \ref{numbercon} we show that given the most general order-$n$ family of densities, we need to impose exactly $n$ conditions to the relative coefficients in order to obtain the most general EGQT density of that order. On the other hand, we have to take into account the existence of the ``trivial'' densities at each order, understood as those that vanish identically for static spherically symmetric metric. As seen above, there are two trivial densities at second order and four at third order. It is, however, simpler to count  first the number of nontrivial densities $\#_{\text{nontrivial}}(n)$. In order to do that, it is convenient to momentarily switch our basis of building-block invariants. Indeed, using the traceless Ricci tensor $\mathcal{S}_{ab}\equiv R_{ab}-\frac{g_{ab}}{3} R$ we consider now
 \begin{equation}
\mathcal{S}_2\equiv  \mathcal{S}_{ab} \mathcal{S}^{ab}\, , \quad  \mathcal{S}_3\equiv  \mathcal{S}_{a}^b \mathcal{S}_{b}^c \mathcal{S}_c^a\, , \quad  \Xi_{1}\equiv \mathcal{S}_{ab}\partial^a\phi \partial^b\phi \, , \quad \Xi_{2}\equiv \mathcal{S}_{ac}\mathcal{S}_b^c\partial^a\phi \partial^b\phi \, ,
\end{equation}
plus $R$ and $\Phi_0$. The benefit of this basis is that it allows one to express the on-shell Lagrangians in a simpler way. Defining the quantities $A\equiv -\left(2f'/r+f''\right)$ and $B\equiv \left(f''-f'/r\right)/3 $, the independent on-shell densities read 
\begin{equation}\label{eq:Sdens}
R|_{\req{eq:SSSM}}=A\, ,\quad  \Xi_{1}|_{\req{eq:SSSM}}=\frac{p^2 B}{r^2} \, , \quad  \mathcal{S}_2|_{\req{eq:SSSM}}=\frac{3B^2}{2}\, , \quad \Xi_{2}|_{\req{eq:SSSM}}=\frac{p^2 B^2}{r^2} \, , \quad  \mathcal{S}_3|_{\req{eq:SSSM}}=\frac{3B^3}{4}\, .
\end{equation}
Now, observe that all of the above, with the single exception of $R$, are  proportional to powers of $B$. Therefore, all the independent densities that we can have at order $n$ in the curvature are $R^i\Xi_{1}^{n-i}$, with $i=0,1,\ldots n$, times arbitrary functions of $\Phi_0$. Thus, the number of nontrivial densities is 
\begin{equation}\label{nontr}
\#_{\text{nontrivial}}(n)=n+1\, .
\end{equation}
This result, combined with the fact that at order $n$ there are $n$ constraints that must be satisfied in order to yield an EGQT density, implies that there is only one nontrivial family of EGQT densities at each order in the curvature ---each family characterized by a function of $\Phi_0$. On the other hand, the number of ``trivial'' densities has a more complicated expression for general $n$. This can be easily obtained as the total number of densities minus the number of densities which belong to the nontrivial set, namely, 
\begin{equation}\label{nontr}
\#_{\text{trivial}}(n)=\#(n)-\#_{\text{nontrivial}}(n)=\#(n)- n-1\, .
\end{equation}
where $\#(n)$ was presented in \req{pin}. Hence, at each order $n$, we have a single nontrivial EGQT and $\#(n)- n-1$ ``trivial''  ones.

In fact, we have been able to obtain an explicit formula for such an order-$n$ EGQT density. It reads
\begin{equation}\label{ggn}
 \mathcal{G}_n=\frac{(-1)^{n-1}}{3^n n}\left[ R+\frac{3\Xi_{1}}{\X}\right]^{n-1}\left[3\left(3nF_{(n)}[\X]+2\X F_{(n)}'[\X] \right)\Xi_{1}+2\X^2 F_{(n)}'[\X] R\right] \, , 
\end{equation}
which of course satisfies the recursive relation \eqref{eq:recurr} and reduces to  \eqref{eq:onshn} when evaluated on-shell on \req{eq:SSSM}. While this reduces to $\mathcal{G}_1$ as in \req{gg1}, the expressions for $\mathcal{G}_2$ and $\mathcal{G}_3$ appearing in \req{gg2} and \req{gg3} differ from the ones corresponding to the $n=2,3$ cases in \req{ggn} by terms which are trivial when evaluated on \req{eq:SSSM}. Observe that $\mathcal{G}_n$ includes terms divided by powers of $\X$ starting at $n=2$. Because of this, if we consider a polynomial expansion for the arbitrary function $F_{(n)}[\X]$, we must demand its lowest-order  term to be $n-1$, \ie 
\begin{equation}
F_{(n)}\left[\X\right]=\sum_{j=n-1}\tilde \beta_{1,j}L^{2(j+n)}\X^{j}\, ,
\end{equation}
 where the $\tilde \beta_{1,j}$ are dimensionless constants. Taking this into account, we can write the most general nontrivial EGQT in three dimensions as
\begin{equation}\label{eq:PLag}
I_{\text{EGQT}}=\frac{1}{16\pi G}\int\diff^3x\sqrt{|g|}\left[R+\frac{2}{L^2}-\sum_{n=0} \mathcal{G}_n\right],
\end{equation}
where $\mathcal{G}_0$ is a general function of $\X$ ---see \req{g00}. Assuming polynomial expansions for the general functions of $\X$ present in the $\mathcal{G}_n$, we have
\begin{align}
\mathcal{G}_0=&+\sum_{k=1}\beta_{0,k} L^{2(k-1)} \X^{k}\, ,\\
\mathcal{G}_1=&-\sum_{k=0} \beta_{1,k} L^{2(k+1)} \X^{ k}\left[(2k+3)\Phi_1- \X R\right]\, ,\\
\mathcal{G}_2=&+\sum_{k=0} \frac{\beta_{2,k }}{2} L^{2(k+3)} \X^{ k}\left[(2k+8)\Phi_1- 2\X R\right]  \Phi_1\, , \\
\mathcal{G}_3=&-\sum_{k=0} \frac{\beta_{3,k }}{3}  L^{2(k+5)} \X^{ k}\left[(2k+13)\Phi_1- 3\X R\right] \Phi_1^2\, , \\ \notag
\dots\\ 
\mathcal{G}_n=&+\sum_{k=0} \frac{(-1)^{n}\beta_{n,k}}{n} L^{2(k+2n-1)} \X^{ k}\left[(2k+5n-2)\Phi_1-n \X R\right]  \Phi_1 ^{n-1}\,,  \quad (n\geq 1)
\end{align}
where in the last line we have written the general form  which includes all  cases with $n\geq 1$.

\subsection{Equations of motion and black holes}
As expected for EGQT gravities, the equations of motion of \eqref{eq:PLag} reduce, when evaluated for static and spherically symmetric solutions, to a single equation for the metric function $f(r)$ which can be integrated once. The resulting equation can be obtained either by direct evaluation of the full non-linear equations of the theory or, alternatively, by considering an ansatz of the form
\begin{equation}\label{gena}
\diff s^2=-N^2(r) f(r)\diff t^2+\frac{\diff r^2}{f(r)}+r^2\diff \varphi^2\, , \quad \phi=\phi(\varphi)\, ,
\end{equation}
$r\in [0,+\infty)$, $\varphi=[0,2\pi)$. Evaluating $L_{N,f,\phi}\equiv \sqrt{|g|} \mathcal{L}_\text{EGQT}|_{\req{gena}}$ and imposing the variations with respect to the three undetermined functions to vanish, 
\begin{equation}
 \frac{\delta L_{N,f,\phi}}{\delta N}=0\, , \quad   \frac{\delta L_{N,f,\phi}}{\delta f}=0\, , \quad  \ \frac{\delta L_{N,f,\phi}}{\delta \phi}=0\, , 
\end{equation}
one finds three equations which are equivalent to the ones obtained from direct substitution of the ansatz \req{gena} on the full non-linear equations ---both for the metric and the scalar--- of the theory \cite{Deser:2003up,Palais:1979rca,Bueno:2021krl}. We find that the second and third equations are proportional to $\diff N/\diff r$ and $\diff^2 \phi/d\varphi^2$, respectively, and  therefore can be solved  by setting $N(r)=N_0$ and $\phi=p \varphi + \phi_0$ where $N_0$ and $\phi_0$ are integration constants which we can set to $1$ and $0$, respectively, without loss of generality. On the other hand, the first equation can be integrated once and the result reads
\begin{equation}\label{EoM}
\frac{r^2}{L^2}-f-\lambda-\beta_{0,1} p^2 \log \frac{r}{L}+\sum_{k=2} \frac{\beta_{0,k}  p^2}{2(k-1)}\left(\frac{p L}{r}\right)^{2(k-1)}+\sum_{n=1}\E_{(n)}=0\, ,
\end{equation}
where $\lambda$ is an integration constant related to the mass of the solution and $\E_{(n)}$ reads
\begin{equation}
\E_{(n)}\equiv \sum_{k=0}\frac{-\beta_{n,k}L^{2(2n-1+k)} p^{2(k+n)} }{n r^{3 n+2k-1}}\left[n(3n+2k-2)f {f'}^{(n-1)}+(n-1)r [ {f'}^{n}-n  f {f'}^{n-2}f'' ]\right]\, .
\end{equation}
As anticipated, the equations of motion depend, at most, on second derivatives of the metric function $f(r)$ for generic higher-curvature theories. Interestingly, it is only for $n=1$ that the expression reduces to an algebraic equation, which means that only at linear order the theory is of the Quasi-topological type. This is of course the set of theories presented in  \cite{Bueno:2021krl} and reviewed in Section \ref{EQTs}.

Although \eqref{EoM} cannot be solved analytically in general, it is possible to establish the existence of black-hole solutions and construct them numerically, as we show below in a few cases. As explained \eg in \cite{Bueno:2016lrh,Bueno:2017qce}, in order for the metric to describe black holes, there are two boundary conditions which need to be satisfied. The first comes from requiring regularity at the horizon, and the other comes from imposing the correct asymptotic behavior, namely, the one corresponding to the Einstein-gravity solutions. Let us analyze these two regimes in more detail.

First, assuming the existence of an outermost horizon, we can consider a Taylor expansion of the solution at $\rh$, namely,
\begin{equation}\label{nearhorizon}
f(r)= 4\pi T (r-\rh)+\sum_{k=2}\frac{a_k}{k!} (r-\rh)^k\, ,
\end{equation}
where, again, we employed the relation $T=f'(\rh)/(4\pi)$, and $a_k\equiv f^{(k)}(\rh)$. 
Substituting this ansatz into (\ref{EoM}), we can solve order by order in the expansion. The first two orders lead to the constraints
\begin{equation}\label{eq:lambn}
\begin{split}
&\frac{\rh^2}{L^2}-\lambda-\beta_{0,1}p^2\log{\frac{\rh}{L}} +\sum_{k=2}\frac{\beta_{0,k}p^2}{2(k-1)}\left(\frac{p L}{\rh}\right)^{2(k-1)} \\ &=\sum_{n=1}\sum_{k=0}\beta_{n,k}\left(\frac{L}{\rh}\right)^{2(2 n+k-1)}\frac{(n-1)}{n}p^{2(k+n)}(4\pi T \rh)^n\, ,
\end{split}
\end{equation}
and
\begin{equation}\label{thermoConstraint}
\begin{split}
&\frac{2\rh}{L^2}-4\pi T- \sum_{k=1}\beta_{0,k}\frac{p^2}{\rh}\left(\frac{p L}{\rh}\right)^{2(k-1)}\\
&=\sum_{n=1}\sum_{k=0}\beta_{n,k}\left(\frac{L}{\rh}\right)^{2(2 n+k-1)}\frac{(3 n+2k-2)}{n \rh}p^{2(k+n)}(4\pi T \rh)^n\, ,
\end{split}
\end{equation}
respectively.
These implicitly relate $\rh$ and $T$ with $\lambda$, $p$ and the gravitational couplings. Meanwhile, going to higher orders shows that all coefficients $a_{n>3}$ are fixed in terms of $a_2$, which turns out to be the only free parameter of the solution. Effectively, $a_2$ will be fixed by requiring the solution to have the correct asymptotic behavior. In order to study that regime, we can assume that the solution takes the form of the charged BTZ solution plus a correction,
\begin{equation}
f(r)\sim \frac{r^2}{L^2}-\lambda-\beta_{0,1} p^2 \log{\frac{r}{L}}+f_p(r)+\epsilon f_h(r)\, .
\end{equation}
Here $f_p(r)$ is a particular solution of the equation \req{EoM} near infinity, obtained as a $1/r$ expansion, 
\begin{equation}\label{fp}
f_p(r)\sim -p^2\left(\frac{p L}{r}\right)^{2\ell}\sum_{n=1}^{\ell+1}\beta_{n,\ell-n+1}2^n\left(\ell+\frac{3}{2}-\frac{1}{n}\right)+\mathcal{O}\left(\frac{\log{r/L}}{r^{2(\ell+1)}}\right)\, ,
\end{equation} 
and $f_h(r)$ represents a perturbation over this solution, controlled by the parameter $\epsilon<<1$. In the above expression we defined $\ell\equiv \text{min} \lbrace n+k-1\mid n>0\rbrace $, that is, the lowest combination $n+k-1$ corresponding to non-vanishing coupling constants $\beta_{n>0,k}$, and we also assumed $\beta_{0,1}\neq 0$. The equation of motion at $\mathcal{O}(\epsilon)$ is an homogeneous differential equation that reads
\begin{equation}\label{homog}
a f_h''+b f_h'+f_h=0\, ,
\end{equation}
where
\begin{equation}
\begin{split}
a&\equiv - \frac{(p L)^{2(\ell+1)}}{r^{2\ell}}\sum_{n=1}^{\ell+1}\beta_{n,\ell-n+1}2^{n-2}(n-1)+\mathcal{O}(r^{-2(\ell+1)}\log{r/L}),\\
b&\equiv +\frac{(p L)^{2(\ell+1)}}{r^{2\ell+1}}(\ell+2)\sum_{n=1}^{\ell+1}\beta_{n,\ell-n+1}2^{n-1}(n-1)+\mathcal{O}(r^{-(2\ell+3)}\log{r/L})\, .
\end{split}
\end{equation}
Note that for $n=1$ we are left with $f_h=0$ and $f(r)\sim \frac{r^2}{L^2}-\lambda-\beta_{0,1} \log{r/L}-p^2 \beta_{1,0}$, in agreement with (\ref{fEQT}). In the general case, the two independent solutions of (\ref{homog}) behave asymptotically as 
\begin{equation}
f_h(r)\sim A r^{\frac{\ell+4}{2}} \exp\left[\frac{r^{\ell+1}}{(\ell+1) (L |p|)^{\ell+1}\sqrt{\rho}}\right]+B r^{\frac{\ell+4}{2}} \exp\left[\frac{-r^{\ell+1}}{(\ell+1) (L |p|)^{\ell+1}\sqrt{\rho}}\right]\, ,
\end{equation}
where 
\begin{equation}
\rho\equiv\sum_{n=1}^{\ell+1}\beta_{n,\ell-n+1} (n-1)2^{n-2}\, .
\end{equation}
This means that regularity at infinity imposes $A=0$, as long as $\rho>0$, and the perturbation remains small (exponentially suppressed) near the asymptotic region. Thus, requiring that the exponential mode is absent at infinity fixes an extra constant and, from the horizon-expansion point of view, this amounts to fixing $a_2$ to a particular value for each set of gravitational couplings. On the other hand, whenever  $\rho<0$ the solutions diverge and both coefficients would need to vanish. Since we have a single available free parameter, we do not expect black-hole solutions to exist for $\rho<0$.    

\begin{figure}
\hspace{-0.2cm}
\begin{subfigure}{.547\textwidth}
  \includegraphics[width=\linewidth]{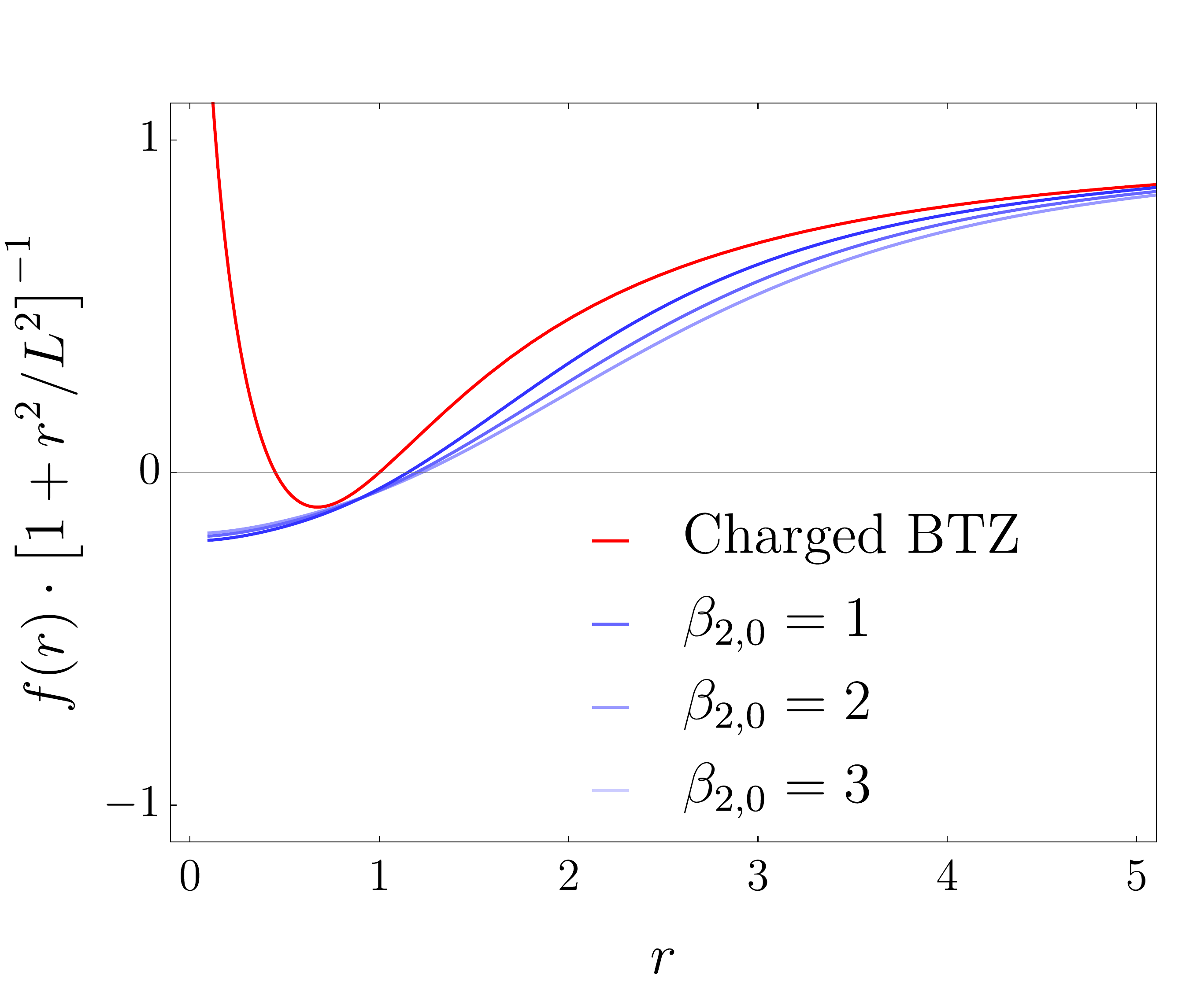}
  \label{fig:sub1}
\end{subfigure}%
\hspace{-1.2cm}
\begin{subfigure}{.547\textwidth}
  \includegraphics[width=\linewidth]{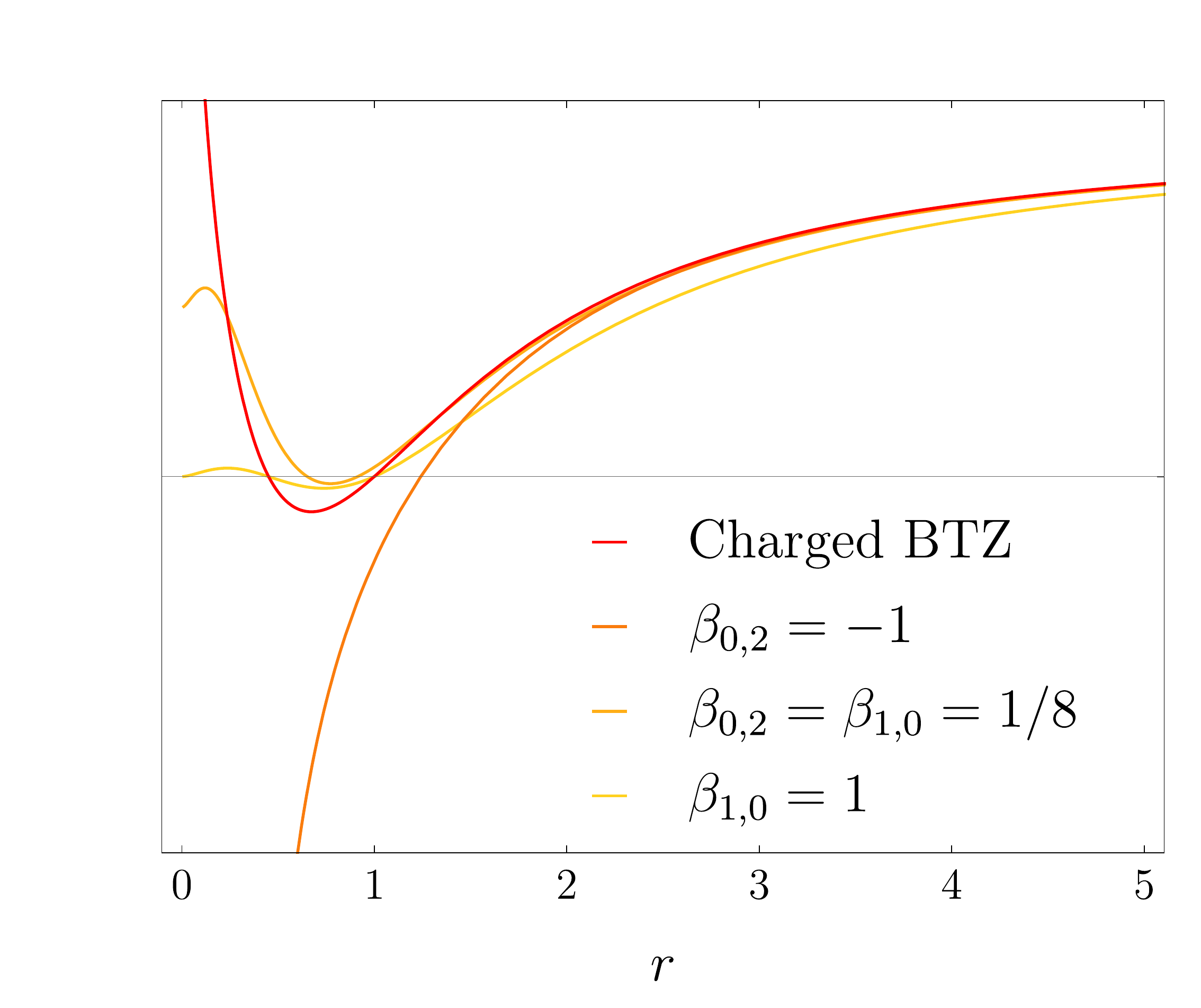}
  \label{fig:sub2}
\end{subfigure}
\caption{Left: We plot the numerical  profiles obtained for black-hole solutions of $n=2$ EGQT theories. We take $L=1$, $p=1$, $\lambda=1$, $\beta_{0,1}=1$, and we set to zero all the rest of couplings, with the exception of $\beta_{2,0}$. Right: We plot profiles corresponding to (analytic) black-hole solutions of EQT gravities. The solutions chosen possess one, two and three horizons, respectively.}
\label{fig:numsol}
\end{figure}

We have solved (\ref{EoM})  numerically in a few cases using the ``shooting method''. That is, we solved for $f(r)$ in a region outside the horizon $r>\rh+\epsilon$, specifying $f(\rh+\epsilon)$ and $f'(\rh+\epsilon)$, according to the expansion (\ref{nearhorizon}). In each case,  we carefully adjusted the coefficient $a_2$ so that the diverging exponential mode at infinity did not pop up and the solution had the right asymptotic behavior. Once $a_2$ was determined, we also solved the differential equation inside the event horizon $r<\rh-\epsilon$, using $f(\rh-\epsilon)$ and $f'(\rh-\epsilon)$ as boundary conditions. Finally we glued the interior and exterior solutions. In the left plot of Fig.\,\ref{fig:numsol} we show some of these solutions, corresponding to $n=2$ EGQT gravities. In each case, we plot the rescaled function $f(r)/[1+r^2/L^2]$, which gives us more control on the solution in the asymptotic region ---in particular, this function is bounded from above when the exponential mode at infinity is absent. The blue curves represent the $n=2$ EGQT black holes, while the red one corresponds to the charged BTZ.  For additional reference, in the right plot we include some of the analytic solutions corresponding to the EQT ($n=1$) theories. As we can see, the new solutions possess a single horizon and they seem to behave near $r\rightarrow 0$ in a way analogous to the neutral BTZ black hole, \ie $f(r) \rightarrow - |\mu|$ for some constant $\mu$. This corresponds to a singularity in the causal structure \cite{Banados:1992gq}. Observe that switching $\beta_{2,0}$ to zero, such solutions reduce to the charged BTZ, which instead has a curvature singularity at the origin. Hence, the introduction of the higher-order coupling tends to smooth out the charged BTZ singularity. It would be interesting to study in more detail the properties of these new black holes, including the possible existence of solutions with additional horizons or other kinds of singularities.



\subsection{Black hole thermodynamics}
In this section we analyze the thermodynamics of the EGQT black holes described above. Although it might be natural to express the thermodynamic quantities as functions of the horizon radius and the charge (or temperature), due to the complicated relation that these satisfy, it proves more convenient to make some redefinitions. We rewrite (\ref{thermoConstraint}) in the form
\begin{equation}\label{constraintThermo}
2-y -\sum_{k=1}\beta_{0,k}x^k-\sum_{n=1}\sum_{k=0}\beta_{n,k} \frac{(3n+2k-2)}{n}x^{n+k}y^n=0
\end{equation} 
where
\begin{equation}
x\equiv \frac{L^2 p^2}{{\rh}^2}\, , \quad \text{and}\quad  y\equiv \frac{4\pi L^2 T}{\rh}\, ,
\end{equation}
 are dimensionless variables. Since expressed in these new variables (\ref{thermoConstraint}) looks tidier, we find it simpler to work with the set $\{\rh,x,y\}$ rather than $\{\rh,T,p\}$. We stress that by virtue of the above equation all the thermodynamic quantities will implicitly depend only on two of these variables, for instance $\{\rh,x\}$.

The entropy, computed using Wald's formula (\ref{Wald}), reads
\begin{equation}
\begin{split}
S
&=\frac{\pi \rh}{2 G}\left[1-\sum_{n=1}\sum_{k=0} \beta_{n,k} x^{n+k} y^{n-1}\right].
\end{split}
\end{equation}
The free energy is calculated according to the procedure described in \cite{Cano:2022ord}, and the result reads 
\begin{align}\notag
 F&=\frac{1}{8G} \left[ \lambda+\beta_{0,1} p^2\log{\frac{r_0}{L}}-2 p^2\beta_{1,0}-\rh f'(\rh)+\sum_{n=1}\sum_{k=0}\beta_{n,k}\frac{L^{2(2n+k-1)} p^{2(k+n)}}{\rh^{3n+2k-2}}f'(\rh)^n \right]\\
&=\frac{\rh^2}{8G L^2} \left[1 +\beta_{0,1} x \log{\frac{r_0}{\rh}}+\sum_{k=2}\frac{\beta_{0,k} }{2(k-1)}x^{k}-2 \beta_{1,0}x-y+\sum_{n=1}\sum_{k=0}\frac{\beta_{n,k}}{n}x^{n+k}y^n\right] \, , \notag
\end{align}
where $r_0$ is an IR cutoff. From the definition $F=M-T S$ one gets for the mass
\begin{align}
 M&=\frac{1}{8G} \left[\lambda+\beta_{0,1} p^2\log{\frac{r_0}{L}}-2 p^2\beta_{1,0}\right]\\ \notag
&=\frac{\rh^2}{8GL^2}\left[1 +\beta_{0,1} x \log{\frac{r_0}{\rh}}+\sum_{k=2}\frac{\beta_{0,k} }{2(k-1)}x^{k}-2 \beta_{1,0}x-\sum_{n=1}\sum_{k=0}\beta_{n,k}\frac{n-1}{n}x^{n+k}y^n\right]\, .
\end{align}
On the other hand, the variable conjugate to the charge is the electric potential $A_t(r)$ at infinity, which is computed from the relation 
\begin{equation}
F_{a b}\equiv 2\partial_{\left[a\right.}A_{\left. b\right]}=4\pi G\epsilon_{a b c}\frac{\partial \mathcal{L}}{\partial(\partial_c\phi)}\, .
\end{equation}
Choosing an integration constant $\Phi$ such that $A_t(\rh)=0$, the solution to the above equation is
\begin{equation}
\begin{split}
A_t(r)=&+\frac{1}{2}\left[\beta_{0,1} p \log{\frac{r}{L}}-\sum_{k=2} \beta_{0,k}\frac{k p}{2(k-1)}\left(\frac{L p}{r}\right)^{2(k-1)}\right.\\
& \left.-\sum_{n=1}\sum_{k=0}\beta_{n,k}\left(\frac{L p}{r}\right)^{2(n+k-1)}\left(\frac{L^2 f'}{r}\right)^n (k+n)\frac{p}{n}+2\beta_{1,0} p\right]+ \Phi
\end{split}
\end{equation}
Hence,
\begin{equation}
\Phi=\frac{\rh \sqrt{x}}{2 L}\left[-\beta_{0,1} \log \frac{\rh}{r_0}-2 \beta_{1,0}+\sum_{k=2}\frac{  \beta_{0,k} k\, x^{k-1} }{2(k-1)}+\sum_{n=1}\sum_{k=0}\frac{\beta_{n,k} (k+n)}{n}x^{n+k-1}y^n \right]\, .
\end{equation}
Finally, written in the suitable variables, the charge and the temperature read 
\begin{equation}
Q=\frac{p}{2 G}=\frac{\rh}{2 G L}\sqrt{x} \, ,
\end{equation}
and
\begin{equation}
T=\frac{y \rh}{4\pi L^2}\, .
\end{equation}
Using (\ref{constraintThermo}), we can check that the first law holds:
\begin{equation}
\frac{\partial M}{\partial \rh}= T \frac{\partial S}{\partial \rh}+\Phi \frac{\partial Q}{\partial \rh}
\end{equation}
and  
\begin{equation}
\frac{\partial M}{\partial x}= T \frac{\partial S}{\partial x}+\Phi \frac{\partial Q}{\partial x}\, .
\end{equation}
In the above expressions we must take into account that all the quantities are implicitly functions of $\rh$ and $x$ alone, so that for a function $f(\rh,x,y)$ we must apply the chain rule  $\frac{\partial f }{\partial x}\equiv\frac{\partial f}{\partial x}+\frac{\partial y}{\partial x}\frac{\partial f}{\partial y}$.

\section{Conclusions}\label{conclu}
In this paper we have characterized and constructed the most general Electromagnetic (Generalized) Quasi-topological theory of gravity in three dimensions. This class includes densities of arbitrarily high curvature orders and generalizes the results presented in \cite{Bueno:2021krl}. Up to terms which make no contribution to the equations of motion when considered for static and spherically symmetric ans\"atze, the most general theory appears in \req{eq:PLag0}. The theories admit solutions of the form \req{eq:SSSM}. The metric function $f(r)$ satisfies a second-order differential equation whenever the EGQT theory includes terms of order $n=2$ or higher in curvatures, and an algebraic one when only $n=0,1$ terms are present ---see \req{eomf}. We have shown that at each curvature order there exists a single EGQT family, $\mathcal{G}_n$, which can be modified by adding to it any linear combination of the $\#(n)-n-1$ order-$n$ densities which are trivial when evaluated on a static and spherically symmetric ansatz ---see \req{pin} for the explicit form of $\#(n)$. We have studied some general properties of the black-hole solutions of these theories, explicitly constructing the profiles numerically in a couple of cases. We have computed the relevant thermodynamic quantities for the most general theory and verified that the first law is satisfied.

It would be interesting to study further the black holes of these theories. In particular, a better understanding of aspects such as the number of horizons or the types of singularities which may arise would clearly be desirable. This would allow for a better comparison with the analytic EQT cases more thoroughly studied in  \cite{Bueno:2021krl}. Unexplored aspects both of the EQT and the new EGQT solutions presented here include the study of their causal structure and orbits, conserved charges, quasinormal modes and dynamical stability ---\eg along the lines of \cite{Hendi:2020yah,Kazempour:2017gho,Gonzalez:2021vwp}. On a different front, EGQT gravities in higher dimensions have been used, in their holographic toy models facet, to prove a universal property satisfied by charged entanglement entropy for general CFTs \cite{Bueno:2022jbl}. In two-dimensional CFTs, the charged R\'enyi entropies display interesting  non-analyticity features in the case of free fields \cite{Belin:2013uta} and it would be interesting to find out if a similar behavior is realized for other theories ---see \eg \cite{Arenas-Henriquez:2022ntz} for work in that direction. Our three-dimensional EGQT gravities  are ideal candidates for this. On a different front, it would be interesting to see if double-copy ideas \cite{Alkac:2021seh,Easson:2022zoh} may make sense for these theories and, if so, characterize the corresponding solutions from the gauge-field perspective.



\section*{Acknowledgements}

We thank Paulina Cabrera, Robie Hennigar and Rodrigo Olea for useful discussions. The work of PB is supported by a Ram\'on y Cajal fellowship (RYC2020-028756-I) from Spain's Ministry of Science and Innovation.
 The work of PAC is supported by a postdoctoral fellowship from the Research Foundation - Flanders (FWO grant 12ZH121N). The work of GvdV is supported by CONICET and UNCuyo, Inst. Balseiro. 
\appendix

\section{Number of conditions for a general-order density to be EGQT  }\label{numbercon}
In this appendix we show that, given a general order-$n$ density, $n$ conditions need to be imposed on the coefficients so that the resulting density is of the EGQT class.
In order to do so, it is convenient to change our basis of invariants. We write now our independent building blocks in terms of the traceless Ricci tensor $\mathcal{S}_{ab}=R_{ab}-\frac{g_{ab}}{3}R$.
Our set of seed invariants reads now
\begin{equation}
\mathcal{S}_2\equiv  \mathcal{S}_{ab} \mathcal{S}^{ab}\, , \quad  \mathcal{S}_3\equiv  \mathcal{S}_{a}^b \mathcal{S}_{b}^c \mathcal{S}_c^a\, , \quad  \Xi_{1}\equiv \mathcal{S}_{ab}\partial^a\phi \partial^b\phi \, , \quad \Xi_{2}\equiv \mathcal{S}_{ac}\mathcal{S}_b^c\partial^a\phi \partial^b\phi \, .
\end{equation}
plus $R$ and $\Phi_0$. The benefit of this basis is that it allows one to express the on-shell Lagrangians in a simpler way. Defining the quantities $A\equiv -\left(2f'/r+f''\right)$ and $B\equiv \left(f''-f'/r\right)/3 $, the independent on-shell densities read 
\begin{equation}\label{eq:Sdens}
R|_{\req{eq:SSSM}}=A\, ,\quad  \Xi_{1}|_{\req{eq:SSSM}}=\frac{p^2 B}{r^2} \, , \quad  \mathcal{S}_2|_{\req{eq:SSSM}}=\frac{3B^2}{2}\, , \quad \Xi_{2}|_{\req{eq:SSSM}}=\frac{p^2 B^2}{r^2} \, , \quad  \mathcal{S}_3|_{\req{eq:SSSM}}=\frac{3B^3}{4}\, .
\end{equation}
In this basis, the general form of an order-$n$ density can be written as
\begin{equation}
\mathcal{L}^{(n)}_{\rm general}=\sum_{j,k,m,p} G_{jkmp,(n)}[\X]R^{n-(2j+3k+m+2p)}\mathcal{S}_2^{j}\mathcal{S}_3^{k}\Xi_{1}^m\Xi_{2}^p \X^{n-m-p}\, ,
\end{equation}
where $G_{jkmp,(n)}[\X]$ are arbitrary functions and the last term is introduced for convenience. The expression of the on-shell Lagrangian evaluated on  \eqref{eq:SSSM} reads
\begin{equation}\label{S_n}
\mathcal{L}^{(n)}_{\rm general}\Big|_{ \eqref{eq:SSSM} }=\sum_{j,k,m,p}\tilde G_{jkmp,(n)}\left[\frac{p^2}{r^2}\right] \left[f''+2\frac{f'}{r}\right]^{n-(2j+3k+m+2p)}\left[f''-\frac{f'}{r}\right]^{(2j+3k+m+2p)},
\end{equation}
where we defined
\begin{equation}
\tilde G_{jkmp,(n)}\left[\frac{p^2}{r^2}\right] \equiv  G_{jkmp,(n)}\left[\frac{p^2}{r^2}\right] \frac{(-1)^{n-(2j+3k+m+2p)}}{18^{j+2k}}\left(\frac{p^{2n}}{r^{2n}}\right)\, .
\end{equation}
Now we follow the steps of \cite{Bueno:2022lhf}, where it was proven, for the case without additional fields besides gravity, that the GQT condition ---namely, the fact that the effective Lagrangian $L_f^{(n)}=\sqrt{-g}\mathcal{L}^{(n)}_{\rm general}\Big|_{ \eqref{eq:SSSM} }$ evaluated on the single-function ansatz becomes a total derivative--- imposes exactly $n$ constraints on the couplings. We will show that the same applies in the EGQT case. We start by employing the binomial identity twice in \eqref{S_n} so that the effective Lagrangian reads
\begin{equation}
L_f^{(n)}=r\sum_{j,k,m,p,q,s}\tilde{\tilde{ G}}_{jkmpqs,(n)}\left[\frac{p^2}{r^2}\right](f'')^{q+s}\left(\frac{f'}{r}\right)^{n-q-s},
\end{equation}
where
\begin{equation}
\tilde{\tilde{ G}}_{jkmpqs,(n)} \equiv \frac{(-1)^{(2j+3k+m+2p)-s}}{2^{2j+3k+m+2p+q-n}}{n-(2j+3k+m+2p)\choose q}{2j+3k+m+2p\choose s}\tilde{ G}_{jkmp,(n)}.
\end{equation}
A necessary condition for $L_f^{(n)}$ to be a total derivative is that the terms with powers of $f''$ higher than one must vanish. Since $0\leq q+s\leq n$, there are $n-1$ constraints that we need to impose in order to remove all such terms and keep the densities of orders $0$ and $1$ in $f''$. By doing so, the general on-shell density becomes 
\begin{equation}\label{lfn}
L_f^{(n)}=r\sum_{j,k,m,p}\left[\tilde{\tilde{ G}}_{jkmp00}\left(\frac{f'}{r}\right)^n+2\tilde{\tilde{ G}}_{jkmp(10)}\left(\frac{f'}{r}\right)^{n-1}f''\right],
\end{equation}
where $2\tilde{\tilde{ G}}_{jkmp(10)}=\tilde{\tilde{ G}}_{jkmp10}+\tilde{\tilde{ G}}_{jkmp01}$. Now, in order for this to be an EGQT density, $L_f^{(n)}$ needs to satisfy the Euler-Lagrangian equation for $f(r)$, which in three dimensions is known to take the form \cite{Bueno:2022lhf}
\begin{equation}
\frac{\partial L_f^{(n)}}{\partial f'}=\frac{\diff}{\diff r}\frac{\partial L_f^{(n)}}{\partial f''}+\text{const.}
\end{equation}
By doing so, we obtain a single additional condition which can be straightforwardly read from \eqref{lfn}. Adding this to the $n-1$ constraints obtained above, we obtain a total of $n$ conditions that one must impose to a general order-$n$ density in order to obtain the most general EGQT density of that order.

\bibliography{Gravities}
\bibliographystyle{JHEP-2}
\label{biblio}

\end{document}